\documentclass[2p,12pt]{elsarticle}
\usepackage[linesnumbered,ruled,vlined]{algorithm2e}
\usepackage{hyperref}
\usepackage[numbers]{natbib}
\usepackage{adjustbox}
\usepackage{graphicx}
\usepackage{booktabs}
\usepackage{xcolor}
\usepackage{lipsum}
\usepackage{amsmath}
\usepackage{amssymb}
\usepackage{array}
\usepackage{soul}
\definecolor{lightblue}{rgb}{0.8,0.85,1}

\usepackage{chngcntr} 

\usepackage[many]{tcolorbox} 
\makeatletter
\def\ps@pprintTitle{%
  \let\@oddhead\@empty
  \let\@evenhead\@empty
  \def\@oddfoot{\reset@font\hfil\thepage\hfil}
  \let\@evenfoot\@oddfoot
}
\makeatother

\begin{document}
\definecolor{yellow}{rgb}{1,1,0} 
\sethlcolor{yellow} 

\begin{frontmatter}


\title{ManuRAG: Multi-modal Retrieval Augmented Generation for Manufacturing Question Answering}

\author[inst1]{Yunqing Li\thanks{Co-first authors}}
\author[inst2]{Zihan Dong\thanks{Co-first authors}}
\author[inst1]{Farhad Ameri\corref{cor1}}
\author[inst3]{Jianbang Zhang}

\affiliation[inst1]{organization={School of Manufacturing Systems and Networks, Arizona State University}, 
            city={Mesa},
            postcode={85212}, 
            state={AZ},
            country={USA}}
            
\affiliation[inst2]{organization={The Department of Mechanical and Aerospace Engineering, North Carolina State University}, 
            city={Raleigh},
            postcode={27607}, 
            state={NC},
            country={USA}}
            
\affiliation[inst3]{organization={AI Lab, Lenovo}, 
            city={Morrisville},
            postcode={27560}, 
            state={NC},
            country={USA}}

\cortext[cor1]{Corresponding author: Farhad Ameri (Farhad.Ameri@asu.edu)}

\begin{abstract}

The evolution of digital manufacturing necessitates intelligent Question Answering (QA) systems that can seamlessly integrate and analyze complex multi-modal data, such as text, images, formulas, and tables. Conventional Retrieval-Augmented Generation (RAG) methods often fall short in handling this complexity, resulting in subpar performance. We introduce ManuRAG, a novel multi-modal RAG framework designed for manufacturing question answering (MQA), incorporating specialized techniques to improve answer accuracy, reliability, and interpret-ability. To benchmark performance, we provide a dataset of 1,515 QA pairs spanning mathematical, multiple-choice, and review questions in manufacturing principles and practices. Experimental results demonstrate that ManuRAG\(_4\) significantly outperforms existing methods across all datasets. Furthermore, ManuRAG’s adaptable design makes it applicable to other domains, including law, healthcare, and finance, establishing it as a versatile tool for domain-specific QA.

\end{abstract}


\end{frontmatter}


\section{Introduction}
\label{sec:intro}
\subsection{Background and Motivation}
In the transition to Industry 5.0, Question Answering (QA) systems play a crucial role in leveraging big data in manufacturing. A QA system is a type of information retrieval system designed to automatically answer questions posed by users in natural language. These systems use a combination of natural language processing (NLP), machine learning (ML), and often knowledge bases or large datasets to understand the question and generate an appropriate answer. By facilitating data-driven decision-making, QA systems provide engineers, operators, and managers with accurate and relevant insights across diverse data formats. This accessibility is essential for intelligent manufacturing, where complex data from sensors, machines, and design files must be integrated smoothly to enable efficient and accurate production.
Integrating information retrieval with systems that prioritize connectivity, automation, and secure collaboration allows manufacturers to streamline the entire value creation process—from ideation and production to sustainment and adaptation. These advanced capabilities promote self-organizing, intelligent production systems and ensure reliable, secure communication across the manufacturing value chain. Additionally, QA systems are crucial for achieving interoperability, enabling seamless interaction between legacy and next-generation equipment and platforms, and for assessing the impact of external factors across a product’s entire life cycle. Potential users of MQA (manufacturing question answering) systems include mechanical design engineers calculating precise tolerances, production engineers optimizing assembly processes, machine operators implementing temperature adjustments, quality assurance teams verifying dimensional accuracy, and research teams exploring advanced materials and techniques for manufacturing efficiency.

\subsection{Challenges}
The complexity of manufacturing data presents a unique challenge for effective information retrieval and QA. Manufacturing documents span a wide array of data types, including text, images, formulas, CAD models, tables, sensor data, and videos~\cite{tao2018data,letaief2020approach,he2023space}. Each format holds critical information necessary for understanding and optimizing manufacturing processes, designs, and operational insights. For example, text data may contain operational guidelines, maintenance logs~\cite{karray2019romain}, or compliance records that require natural language processing (NLP) for efficient retrieval. Images, like defect photos or assembly line snapshots, demand computer vision techniques to identify patterns or detect quality issues. Formulas and equations, embedded in documentation, need precise recognition and interpretation to enable accurate replication and troubleshooting of complex processes. Together, these varied data formats underscore the essential role of advanced information extraction and processing techniques for efficient, data-driven manufacturing insights.

The adoption of MQA systems in industry remains in its early stages, driven by the growing complexity of manufacturing processes and the increasing need for advanced data management solutions. GPT-powered chatbots~\cite{chatgpt2025}, such as those offered by Streebo~\cite{streebo_chatgpt_manufacturing}, represent the closest analogs. However, these tools primarily focus on conversational interactions, lacking the robust multi-modal retrieval and domain-specific reasoning capabilities required for comprehensive MQA applications.

Traditional information retrieval methods for MQAs are typically limited to either text-based or image-based data retrieval, often failing to capture the full context and complexities of multi-modal information. This limitation can lead to inefficiencies in production workflows and hinder effective decision-making. For instance, in design and development departments—especially within complex industries like automotive manufacturing—retrieving relevant information across various data formats presents significant challenges. An engineer, for example, might need to answer, “Among the following models, which one has a five-link suspension? Options: A: Model 1, B: Model 2, C: Model 3.” The answer could be spread across multiple document types, where older models may store information as text, Model 2’s data might be embedded in a PDF with drawings and technical specifications, and Model 3’s details could be in CAD models, exploded views, or annotated images. Manually sifting through these formats is time-intensive, increases the risk of missing key details, and becomes even more difficult with extensive documentation. This example highlights a core issue: the pressing need for a robust retrieval system that can seamlessly integrate and interpret text, visual data, and structured tables in a way that meets the complex needs of design and manufacturing.

Several factors limit today’s information retrieval systems for QA in the manufacturing domain. Manufacturing data is diverse, intricate, and technical, requiring specialized processing for each type~\cite{STARLY202350,li2024manufacturing,bharadwaj2022knowledge}. Textual data involves domain-specific terminology, while visual data like figures, tables, and slides demand advanced interpretation tools—capabilities lacking in many current systems. Another challenge is transforming text and image data into a unified format computers can process, with minimal information loss~\cite{lu2024ovis}. These issues lead to misinterpretation or loss of critical details, hindering efficient data retrieval and integration. Advanced multi-modal retrieval approaches are essential to enable faster, accurate decision-making in modern manufacturing.

\subsection{Objectives}

To overcome these limitations, we introduce a multi-modal retrieval-augmented generation (MMRAG) system tailored for MQA. This system seamlessly integrates and processes various data types—such as images, formulas, tables, and text—to deliver a deeper and more contextually accurate understanding of manufacturing documents. By bridging the gap between diverse data formats and complex manufacturing queries, MMRAG empowers users with actionable insights that are both reliable and comprehensive. It facilitates more informed decision-making, streamlines workflows, improves operational efficiency, and minimizes production errors, ultimately enhancing productivity and driving innovation in manufacturing systems.

This paper seeks to leverage the power of multi-modal data integration and retrieval to enhance MQA and decision support. The main contributions of this paper include:

\begin{enumerate}
\item Introduces a comprehensive multi-modal QA system tailored to the manufacturing domain, designed to identify operational bottlenecks, improve product design, and support data-driven decision-making in complex manufacturing processes.

\item Designs and evaluates various QA systems, including traditional RAG and MMRAG frameworks, offering valuable insights into diverse manufacturing scenarios while extending their applicability to other domains such as law, healthcare, and finance.

\item Proposes an innovative approach to multi-modal information indexing and retrieval by utilizing links embedded in textual data and aligning text with images either before or after embedding, enabling a more robust and accurate interpretation of visual information to enhance MQA system performance.
\end{enumerate}

The remainder of the paper is organized as follows: in Section~\ref{sec:relate}, we review related work. Section~\ref{method} introduces how we construct ManuRAG. In Section~\ref{sec:exp}, we present experiments demonstrating the effectiveness of our approach. Finally, Section~\ref{sec:con} discusses the limitations of our method and suggests directions for future work.

\section{Related Work}
\label{sec:relate}


\subsection{Question Answering Systems}

QA systems have evolved significantly, leveraging a range of techniques to retrieve and respond to queries~\cite{soares2020literature}. Traditional expert systems, built on domain-specific expertise, have long been used in QA tasks but often lack scalability and adaptability for complex, data-rich environments. Large language models (LLMs) such as GPT~\cite{achiam2023GPT} and BERT~\cite{vaswani2017attention} have introduced major advancements in QA by using deep learning to interpret queries and retrieve relevant information from extensive datasets. These models excel in understanding context and language, making them highly effective in general-purpose QA. Multi-modal models have further advanced the field, enabling visual question answering (VQA) by combining image and text processing~\cite{singh2019towards} and supporting QA with the integration of tabular and textual data for contextually rich responses~\cite{zhu2021tat}. However, none of these models are suitable for manufacturing domain due to the complexity and highly technical nature of manufacturing data. Manufacturing question answering(MQA) requires precise extraction and robust integration of diverse, multi-modal data types which presents a unique challenge beyond the capabilities of existing QA models.

\subsection{Information Extraction}
Information extraction (IE) has evolved to handle diverse formats in structured and unstructured data, including text, images, tables, and multimedia. Multimedia documents, like PDFs, pose unique challenges by integrating text, images, tables, and sometimes audio or video. Advanced IE approaches use layout detection models such as R-CNN~\cite{soto2019visual} and LayoutLMv3~\cite{huang2022layoutlmv3} to identify content regions. Formula recognition relies on deep learning methods~\cite{gao2017deep, zhou2022end, terven2023comprehensive, wang2024unimernet}. OCR tools~\cite{ma2019paddlepaddle,chaudhuri2017optical} are essential for accurate text extraction across fonts and languages, providing comprehensive IE capabilities for multimedia content. IE in manufacturing involves extracting information from text, images, and other data types to support decision-making. For text, Li et al.\cite{li2024building} developed methods to extract knowledge from manufacturer websites, streamlining service discovery. For images, Bhandarkar and Nagi\cite{bhandarkar2000step} convert geometric data into higher-level manufacturing insights. Despite the prevalence of multimedia in manufacturing, no studies have focused on extracting information from these diverse sources, highlighting a research gap.

\subsection{Prompt Engineering and Retrieval Augmented Generation}

Large Language Models (LLMs) perform well in general QA but face challenges in manufacturing due to specialized terminology, complex data, and technical detail needs ~\cite{xue2024domain}. Manufacturing queries require precise, domain-specific knowledge that LLMs, prone to generating plausible but incorrect answers (hallucinations)~\cite{yao2023llm}, cannot reliably provide without specialized grounding, making them insufficient for MQA without further refinement.

Prompt engineering involves crafting prompts to guide LLMs toward accurate, contextually relevant responses. Techniques like Chain of Thought~\cite{wei2022chain} and Tree of Thought~\cite{yao2024tree}, enhance this process by promoting multi-step reasoning, enabling LLMs to handle complex queries more effectively. In MQA, prompt engineering integrates industry-specific language and technical constraints into tailored prompts, enabling LLMs to produce precise, domain-relevant answers and effectively manage specialized terminology and complexity~\cite{xia2024leveraging,valentine2024prompt}.

Another powerful approach is RAG, which grounds LLM outputs in external knowledge by retrieving relevant documents or databases from real-world manufacturing data. Tools like Llama-Index\cite{llamaindex} and LangChain\cite{langchain} enable efficient retrieval, while GraphRAG~\cite{graphrag} introduces graph-based retrieval for reasoning over interconnected data. RAG enhances LLM outputs by retrieving relevant manufacturing data from documents or databases. Methods like text embedding RAG and graph embedding RAG match queries with documents or capture relationships within interconnected data. Tools like Llama-Index\cite{llamaindex} and LangChain\cite{langchain} support text embedding, while GraphRAG\cite{graphrag} uses graph-based retrieval for complex reasoning. In manufacturing, RAG applications range from retrieving domain-specific terms~\cite{bei2024manufacturing}, integrating ontological knowledge graphs for engineering analysis, to materials design~\cite{buehler2024generative}. Multimodal RAG, an emerging extension that integrates text, image, and other data types to provide a richer context for complex questions~\cite{sun2024multimodal,zhu2024realm}, has shown applications in healthcare but has not been tailored for the manufacturing domain.

Multi-modal large language models (MLLMs), such as Ovis~\cite{lu2024ovis}, GPT-4O, and Qwen~\cite{qwen2.5}, have significantly advanced the integration of text and visual data. While these MLLMs excel in understanding multi-modal inputs, they are constrained by input token limits, which pose challenges when processing massive amounts of information. 

MMRAG addresses this limitation by indexing large-scale information and retrieving relevant context during inference, enabling efficient access to extensive external knowledge. Unlike traditional RAG, which focuses on text-based data, MMRAG extends this capability to incorporate both text and visual modalities. MMRAG typically employs two primary approaches to handle diverse data formats: (1) converting images into descriptive text for seamless integration into text-based workflows, and (2) aligning text and image data in a shared embedding space using multi-modal embeddings like CLIP~\cite{radford2021learning} and Jina-CLIP~\cite{xiao2024jina}. These methods allow models to retrieve and reason across modalities, enhancing their performance on tasks that require both visual and textual comprehension. While multimodal RAG has shown applications in domains such as healthcare, its potential remains underexplored in manufacturing. The manufacturing domain demands tailored solutions capable of handling intricate schematics, technical diagrams, and process-specific content with high precision and contextual relevance.

\section{Methodology}
\label{method}

Traditional RAG systems are primarily designed to handle text inputs, often overlooking the multi-modal nature of information present in real-world data. To overcome this limitation, we propose ManuRAG, an extension of RAG that incorporates multimodality to process diverse data types from manufacturing documents. The MLLM employed in ManuRAG accepts both plain text and images as inputs, enabling it to process \textit{.pdf} files (PDFs) through a pipeline of object detection and text recognition models. This process effectively categorizes documents into distinct components such as figures, images, formulas, and textual content.

The ManuRAG framework comprises four key stages: multimodal data extraction, indexing, retrieval, and answer generation. A state-of-the-art PDF extraction pipeline—integrating advanced vision and segmentation models—facilitates robust figure and content recognition \cite{wang2024mineruopensourcesolutionprecise}. Details of the multimodal data extraction process are presented in Section \ref{data-extraction}. Indexing mechanisms are discussed in Section \ref{index}, retrieval strategies in Section \ref{retrieval}, and answer generation approaches in Section \ref{generation}.

Additionally, we introduce four variations of the ManuRAG framework. These variations explore different strategies for integrating multi-modal data. The details of these variations are presented in Section \ref{variation}.

\subsection{Multi-modal Data Extraction}
\label{data-extraction}

Using original manufacturing files directly in RAG presents significant challenges. These files often contain complex structures, including images, tables, formulas, and non-standard text layouts, which make direct use infeasible for efficient information retrieval. Converting PDFs to plain text frequently results in the loss of crucial contextual and structural information, yielding low-quality data that undermines the performance of MQA To address these challenges, we focus on extracting and organizing multi-modal data—text, images, tables, and formulas—from manufacturing PDFs to create a robust input for RAG systems.

The primary goal of this section is to construct high-quality data representations that accurately capture fundamental manufacturing knowledge while minimizing noise. This process begins with extracting and processing information from manufacturing PDF files to create separate text and image representations, ensuring the preservation of critical data. Using the pdf-extract-kit framework\cite{wang2024mineru,zhao2024doclayoutyoloenhancingdocumentlayout,wang2024unimernet,he2024opendatalab}, we employed state-of-the-art methodologies for multi-modal data extraction. This includes layout detection to maintain the document's structure, formula extraction to retain mathematical content, and image and table extraction to incorporate visual data. Optical character recognition (OCR) was applied to digitize textual content from non-selectable regions. These multi-modal representations are then embedded and stored in a unified database that preserves relationships between different modalities, enabling seamless integration and effective use in downstream tasks.

\subsubsection{Layout Detection}

Accurate layout detection is crucial for maintaining the structural integrity of manufacturing PDFs and extracting relevant content effectively. This process entails parsing documents into distinct elements while filtering out irrelevant components, such as page numbers or titles, that could negatively impact QA performance. To achieve this, we employed LayoutLMv3~\cite{huang2022layoutlmv3} to detect regions like images, tables, titles, and text, and 

\subsubsection{Formula Extraction}

We employ YOLOv8~\cite{hussain2023yolo} to identify inline and block formulas with precise spatial locations. This method is particularly effective at identifying inline formulas, such as $E = \frac{\sigma}{\epsilon}$ (Young’s modulus formula, where $\sigma$ is stress and $\epsilon$ is strain), or $J = -D \frac{\partial C}{\partial x}$ (Fick’s first law, where $J$ is diffusion flux, $D$ is the diffusion coefficient, and $\frac{\partial C}{\partial x}$ is the concentration gradient). By distinguishing formulas from surrounding textual content, the detection method ensures the preservation of critical mathematical expressions, preventing errors during text extraction and maintaining the contextual integrity of manufacturing documents.

\subsubsection{Image and Table Extraction}
Images and tables are indispensable tools for conveying dense information intuitively and are widely used in manufacturing systems. However, their complex structure poses challenges for integration into QA systems. To address this, we implemented a pipeline to convert tables into LaTeX formats, using a converter model trained on the PubTabNet dataset ~\cite{zhong2019image}, ensuring accurate and structured tabular data extraction. 

\subsubsection{Optical character recognition}
After processing non-textual elements, plain text was extracted using Paddle-OCR~\cite{paddleocr}. This approach avoids errors caused by processing entire pages as single inputs. Extracted content is stored with metadata that included location and type information: formulas in LaTeX, figures and tables as image files, and plain text in markdown format. Each visual element is assigned a Universally Unique Identifier (UUID), a 128-bit identifier used to reference these elements consistently within the text, enabling seamless integration. This structured methodology ensures a high-quality and well-organized representation of the data.

\label{index}
\subsection{Indexing}

Indexing plays a crucial role in RAG, as it facilitates efficient and accurate retrieval of relevant information during query resolution. In ManuRAG, indexing involves structuring and storing multi-modal data, including both textual and visual information, to enable seamless integration into subsequent retrieval and generation tasks. By projecting data into a high-dimensional embedding space, the system ensures compatibility with similarity-based searches, which are essential for effective information retrieval.

\subsubsection{Text Vector Database}
Textual data, including paragraphs, formulas and links to the images, is pre-processed dividing it into manageable chunks to preserve contextual relevance and coherence. This segmentation process is governed by two key parameters: the chunk size, which defines the maximum length of each segment, and the chunk overlap, which determines the degree of overlap between consecutive chunks to maintain continuity of context. Given an textual input \textit{Text}, the division into overlapping chunks can be expressed as:

\begin{equation}
\mathrm{Chunks}
\;=\;
\bigl\{\,C_i : C_i = \textit{Text}[\,i : i + S + O\,],\; i \in [0,\,|\textit{Text}|],\;\text{step}=S\,\bigr\}
\label{eq:chunking}
\end{equation}

where \( S \) represents the chunk size, \( O \) denotes the chunk overlap, \( C_i \) is the \( i \)-th chunk, and \( |T| \) is the total length of the text. 

We model each text chunk $C_i$ as a pair:
\begin{equation}
C_i = \bigl(T_i, \mathcal{\alpha}_i \bigr),
\quad
\mathcal{\alpha}_i = \{\, I_{i1}, I_{i2}, \dots, I_{i\,b_i} \},
\label{eq:chunk_model}
\end{equation}
where $T_i$ is the textual content, and $\mathcal{\alpha}_i$ is the set of images referenced in $T_i$

The chunks are projected into the embedding space using a LLM-based text embedding model \( \text{Embed}_{\text{Text}} \). The embedding for each chunk \( C_i \) is represented as:

\begin{equation}
v_{C_i} = \text{Embed}_{\text{Text}}(C_i)
\label{eq:text_embedding}
\end{equation}

These embeddings, along with their corresponding text chunks, are stored in a text vector database, enabling efficient retrieval during the query resolution process.

\subsubsection{Image Vector Database}

For visual data, including extracted figures and tables, a multi-modal embedding model is employed to encode the information into the same embedding space as the text. This model is capable of handling both text and images, thus enabling cross-modal search capabilities, such as text-to-image and image-to-text retrieval. Each image \( I_j \) is encoded in an embedding vector \( E_{I_j} \) using the multi-modal embedding function \( \text{Embed}_{\text{MM}} \), as defined by:

\begin{equation}
v_{I_j} = \text{Embed}_{\text{MM}}(I_j)
\label{eq:image_embedding}
\end{equation}

The resulting image embeddings, along with their associated image data, are stored in an image vector database. This database is designed to support the retrieval of visual information and to complement the textual data during queries.

Although the multi-modal embedding model is capable of processing both text and images, its understanding of complex and lengthy textual content is not as good as that of most LLM-based text embedding models. The text embedding model, which is trained specifically on extensive textual datasets, provides a more comprehensive understanding of intricate text structures and semantics. To capitalize on the strengths of both approaches, some variations of ManuRAG maintain separate vector databases for text and images. The details are explained in Section \ref{variation}. This dual-database approach allows ManuRAG to effectively leverage the specialized capabilities of the text embedding model and the flexibility of the multi-modal embedding model.

\subsection{Retrieval}
\label{retrieval}

Retrieval is a key component of RAG system, enabling the system to identify and extract the most relevant information from the indexed data. In ManuRAG, retrieval serves the critical function of narrowing down the multimodal data—text and images—stored in vector databases, so that only the most pertinent chunks are provided as context to the language model for generating answers to manufacturing-related questions. By incorporating retrieval, the system ensures efficiency and accuracy in handling large and diverse datasets.

The retrieval process is based on calculating the similarity between the embedding of an input manufacturing query and the embeddings of stored data. The similarity metric used is cosine similarity, defined as:

\begin{equation}
\text{Sim}(\mathbf{q}, \mathbf{v}) = \frac{\mathbf{q} \cdot \mathbf{v}}{\|\mathbf{q}\| \|\mathbf{v}\|}
\label{eq:cosine_similarity}
\end{equation}

where \( \mathbf{q} \) is the embedding of the input manufacturing question  \( \mathbf{Q} \) , and \( \mathbf{v} \) is every embedding vector of a stored data vector (text chunk or image). Cosine similarity evaluates the angular alignment between the query and the stored embeddings, providing a measure of relevance.

\subsubsection{Retrieval from the Text Vector Database}

For textual data, the system calculates cosine similarity between the query embedding and the embeddings of all indexed text chunks. Using cosine similarity $\mathrm{Sim}$ from Eq.~\eqref{eq:cosine_similarity}, the top-$k$ text chunks are retrieved according to:

\begin{equation}
Top@K_{text}= \text{argmax}_k \big(\text{sim}(\mathbf{q_1}, v_{C_i})), \forall i
\end{equation}

where $\mathbf{q}_1$ is the query embedding in the text space, and $\mathbf{v}_{C_i}$ is the embedding of chunk $C_i$. 
Once retrieved, we collect the set of all images referenced by these chunks:
\begin{equation}
\mathrm{Image_{referenced}}
= \mathcal{\alpha}_{i_r} =\, \bigcup_{r=1}^k \bigl\{\, I_{\,i_r 1}, \, I_{\,i_r 2}, \,\dots\bigr\}.
\label{eq:collected_images}
\end{equation}

\subsubsection{Retrieval from the Image Vector Database}

For visual data, the input query is encoded into the same embedding space as the image embeddings using a multi-modal embedding model. Cosine similarity is then calculated between the query embedding and the image embeddings stored in the image vector database. The top \( k \) images with the highest similarity scores are retrieved:

\begin{equation}
Top@K_{Image}= \text{argmax}_k \big(\text{sim}(\mathbf{q_2}, v_{I_j})), \forall j
\end{equation}

where \(q_{2}\) is the embedding of the input query in the multi-modal embedding space, \(v_{I_j}\) represents the embedding of the \( j \)-th image \(I_j\). This enables the relevant visual data to be directly retrieved by embedding similarity.

\subsection{Answer Generation}
\label{generation}

Answer generation is the final stage of RAG pipeline, where the system synthesizes a response to the input query by leveraging the relevant data retrieved during the retrieval phase. This step involves integrating the retrieved textual and visual information with the manufacturing question and processing it through an LLM or MLLM to produce an accurate and contextually appropriate answer.

\subsubsection{Answer Generation Using Text-Only Retrieval Data}

In the first strategy, the system combines the retrieved textual data, the manufacturing question, and the domain-specific prompt as input to a large language model (LLM) optimized for text-based processing. This approach excludes any retrieved visual data, focusing solely on textual context for answer synthesis. The process can be expressed as:

\begin{equation}
\text{A}_{\text{LLM}} = \text{LLM}(\text{concat}(\text{P}, Q, Top@K_{text}))
\label{eq:text_only_strategy_with_prompt}
\end{equation}

where \( P \) is the prompt specially designed manufacturing domain questions, \( Q \) represents the input query. This strategy is efficient and suitable for cases where the retrieved textual data alone provides sufficient context to answer the query. However, it does not account for visual information, which may limit its effectiveness for multi-modal queries.

\subsubsection{Answer Generation Using Multi-modal Retrieval Data}

In the second strategy, the system combines the retrieved textual data, retrieved visual data, and the manufacturing question as input to a multi-modal language model (MLLM). This approach incorporates both text and images, enabling the model to consider visual context alongside textual information for answer generation. The answer generated can be formulated as:

\begin{equation}
\text{A}_{\text{MLLM}} = \text{MLLM}(\text{concat}(P,Q, Top@K_{text}, Top@K_{Image},\mathcal{\alpha}_{i_r}))
\label{eq:multimodal_strategy}
\end{equation}

This strategy generates answers using relevant images retrieved both through images referenced in the text and directly from the image vector database. It is important to note that \( Top@K_{\text{Image}} \) and \( \mathcal{\alpha}_{i_r} \) may not always exist in cases where no visual data is linked, indexed or retrieved.

We include our designed prompts in \ref{appendix:Prompt} for different types of manufacturing questions and we will give more introduction to the difference between ManuRAG variations in Section~\ref{variation} below.

\subsection{ManuRAG Variations}
\label{variation}

\begin{figure}[!h]
    \centering
    \includegraphics[width=0.99\linewidth]{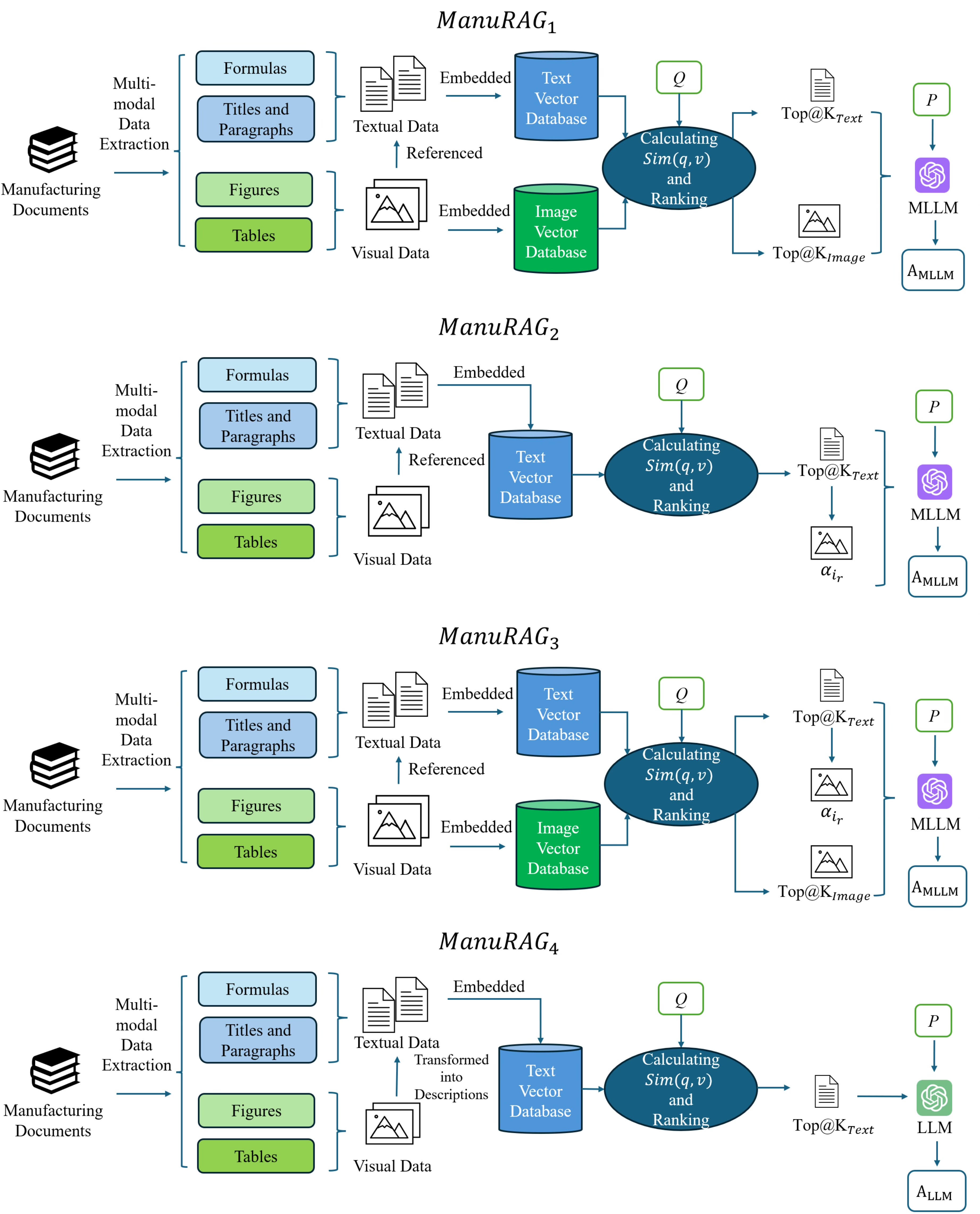}
    \caption{Four Types of ManuRAG Frameworks}
   \label{fig:manurag_variations}
\end{figure}
\label{sec:va}

We propose four variations of the ManuRAG framework to investigate the optimal approach for aligning textual and visual data in MQA. Specifically, the framework explores different strategies for integrating text and image data—whether alignment should occur before embedding, within the embedding space, or after embedding. Figure \ref{fig:manurag_variations} highlights the workflows of each variation, emphasizing the differences in multi-modal data handling, indexing strategies, and answer generation methods.

ManuRAG$_1$ processes both textual and visual data by embedding textual information into a text vector database and visual data into an image vector database. Answer generation is performed using an MLLM.

ManuRAG$_3$ extends ManuRAG$_1$ by prioritizing the retrieval of visual data linked within the retrieved text chunks. It uses both text and image vector databases for indexing but focuses on retrieving images referenced in the text.

ManuRAG$_2$ simplifies ManuRAG$_3$ by eliminating the image vector database. It operates exclusively on textual data, storing embeddings in a text vector database, and generates answers using a LLM. 

ManuRAG$_4$ introduces a unique step in multi-modal data processing. After identifying visual data referenced in the text, the visual content (e.g., figures and tables) is transformed into descriptive text using an MLLM. This transformed text is then indexed in the text vector database, and answers are generated using an LLM (or optionally an MLLM). This variation aligns textual and visual data before embedding by converting visual data into a text-based representation. However, this transformation may result in some information loss, especially when the original visual context is critical.








\section{Experiments}
\label{sec:exp}
In this section, we present experiments to evaluate the effectiveness of the proposed methods for MQA using real-world datasets to ensure practical relevance. The study focuses on evaluating how different RAG systems improve the performance of MQA and examining the influence of context precision and context recall on the effectiveness of these systems in retrieving relevant information. The experimental settings, including datasets, baselines, and configurations, are detailed in Section \ref{4.1}, while the evaluation metrics are introduced in Section \ref{metrics}. The general performance of the models and the context-retrieval-related performance are discussed in Sections \ref{4.3} and \ref{4.4}, respectively.

\subsection{Settings} 
\label{4.1}  

\subsubsection{Baselines and Configurations}
We evaluate multiple variations of ManuRAG against alternative QA solutions to demonstrate the performance improvements and unique advantages offered by our approach. All models in the comparison leverage the same underlying MLLM, GPT-4O, ensuring a consistent foundation for multi-modal capabilities. Our evaluation encompasses the baseline GPT-4O, both in its standard configuration and with manufacturing-specific prompts, as well as various RAG models, including traditional and multi-modal RAG approaches.

All RAG models employ Ada v2 for text indexing and retrieval, CLIP for image indexing and retrieval when an image vector store is utilized, and GPT-4O for generating final answers. Both the ManuRAG variants and the RAG methods are built using the LlamaIndex framework~\cite{llamaindex}. The source documents for RAG context retrieval are derived from the same PDF version of a textbook, \texttt{Fundamentals of Modern Manufacturing}~\cite{groover2010fundamentals}, which is a widely recognized textbook covering engineering materials, manufacturing processes, and electronics technologies.

All experiments are run in Google Colab with \(k=1\), \(S=600\), and \(O=100\), following Edge et al. The values of \(k\), \(S\), and \(O\) are selected based on their performance in a 5\% sample test. Except for the baseline models, all other models utilize a unified prompt configuration, as detailed in~\ref{appendix:Prompt}. Below are additional details about the benchmark models:

\begin{itemize}
    \item \textbf{GPT-4O}: Direct QA using GPT-4O without any specialized prompts.

    \item \textbf{GPT-4O (CoT)}: Direct QA enhanced with Chain-of-Thought (CoT) prompting to enable step-by-step reasoning with GPT-4O.

    \item \textbf{RAG}: RAG applied to raw text directly extracted from the files, excluding any visual data.

    \item \textbf{RAG\(_{hq}\)}: A high-quality RAG framework utilizing highly structured text extracted using the methodology outlined in Section~\ref{data-extraction}.

    \item \textbf{ManuRAG\(_{1-3}\)}: Variants of MMRAG, as described in Section~\ref{variation}.

    \item \textbf{ManuRAG\(_{4}\)}: A variant of MMRAG introduced in Section~\ref{variation}, incorporating GPT-4O to transform image data into text for enriched contextual understanding.
\end{itemize}

\subsubsection{Evaluation Datasets}

We evaluate our benchmarks using three datasets consisting of 1,515 question-answer pairs derived from the official solutions of Fundamentals of Modern Manufacturing~\cite{groover2010fundamentals}. Notably, this dataset is excluded from the content used for RAG retrieval. The datasets are organized as follows:

\begin{itemize}
    \item \textbf{MathQ}: 309 mathematical QA pairs requiring factual accuracy and semantic understanding, often multi-modal with formulas, tables, and figures. Preprocessing excludes excessively long questions/answers (over 600 tokens) or trims sub-questions for LLM compatibility.
    
    \item \textbf{MCQ}: 471 multiple-choice QA pairs ssessing the ability to select correct options, with some requiring multi-modal analysis. Answers are preprocessed into distinct choices (e.g., abcd) with explanations to ensure precise evaluation.

    \item \textbf{RQ}: 735 review QA pairs focusing on comprehension and reasoning over detailed textual content to evaluate conceptual understanding.
\end{itemize}

Examples of each type of datasets are provided in ~\ref{appendix:QA}.

\subsection{Evaluation Metrics}
\label{metrics}

We adopt five evaluation metrics from RAGAS\cite{es2023ragas} with GPT-4 to comprehensively assess the performance of QA models which are Factual Correctness, Semantic Semilarity, ROUGE, Context Precision and Context Recall. The detailed metrics and their formulas are as follows.

\subsubsection{Factual Correctness}

Factual correctness measures the F1 score of the generated answers by comparing them to the ground truth and is applied across all evaluated models. This metric assesses the degree to which the generated response aligns with the reference, with scores ranging from 0 to 1, where higher values indicate better performance. GPT-4 is employed to decompose both the response and the reference into distinct claims, with natural language inference used to evaluate the factual overlap between them. By balancing precision and recall through the F1 score, this metric provides a comprehensive measure of the accuracy of the generated answers relative to the reference.

True Positive (TP) represents the number of claims in the answer that are present in the reference. False Positive (FP) indicates the number of claims in the response that are not present in the reference, while False Negative (FN) corresponds to the number of claims in the reference that are not present in the answer. Using these values, the precision, recall, and F1 score are calculated as follows:

\begin{equation} \text{Precision} = \frac{\text{TP}}{\text{TP} + \text{FP}} \tag{1} \end{equation}

\begin{equation} \text{Recall} = \frac{\text{TP}}{\text{TP} + \text{FN}} \tag{2} \end{equation}

\begin{equation} \text{Factual Correctness (F1 Score)} = \frac{2 \times \text{Precision} \times \text{Recall}}{\text{Precision} + \text{Recall}} \tag{3} \end{equation}

\subsubsection{Semantic Similarity}

Semantic similarity evaluates how semantically close the generated answer is to the reference answer, which is is applied across all evaluated models. It is measured using a cross-encoder model that computes a semantic similarity score ranging from 0 to 1. A higher score indicates a better semantic alignment between the generated answer and the ground truth.

\subsubsection{ROUGE}

ROUGE evaluates the overlap between the generated answers and the reference answers in multiple-choice questions (MCQ). It measures the overlap of \(n\)-grams (\(n=1\)) using the F1 score to quantify the similarity between the generated response and the reference. The formula for ROUGE is defined as follows:

\begin{equation}
\text{Precision}_n = \frac{\text{Number of overlapping \(n\)-grams}}{\text{Total number of \(n\)-grams in the generated answer}}
\end{equation}

\begin{equation}
\text{Recall}_n = \frac{\text{Number of overlapping \(n\)-grams}}{\text{Total number of \(n\)-grams in the reference}}
\end{equation}

\begin{equation}
\text{ROUGE}_n = \frac{2 \times \text{Precision}_n \times \text{Recall}_n}{\text{Precision}_n + \text{Recall}_n}
\end{equation}

This metric provides a straightforward and effective way to assess lexical similarity in MCQ dataset, where exact matches of words or phrases are crucial for evaluation.

\subsubsection{Context Precision}
Context precision measures the relevance of retrieved context chunks to the query. It evaluates the proportion of relevant chunks among the retrieved contexts in RAG-based models, including RAG\(_1\), RAG\(_2\), and ManuRAG\(_1\) to ManuRAG\(_4\). This metric incorporates a weighted precision approach, where \(v_k\) represents the weight assigned to each chunk at rank \(k\). The precision at rank \(k\) (\texttt{$top_k=3$}) is computed as:

\begin{equation}
\text{Precision@k} = \frac{\text{True Positives@k}}{\text{True Positives@k} + \text{False Positives@k}} \tag{1}
\end{equation}

Using these precision values, the context precision for the top \(K\) results is calculated as:

\begin{equation}
\text{Context Precision@K} = \frac{\sum_{k=1}^{K} (\text{Precision@k} \times v_k)}{\text{Total number of relevant items in Top@K}} \tag{2}
\end{equation}

\subsubsection{Context Recall}

Context recall measures the percentage of relevant context retrieved compared to the total relevant context available among the retrieved contexts in RAG-based models. Ground Truth (GT) refers to the set of reference claims or facts that are known to be correct and are used as a standard for evaluation. This metric calculates how many of the claims in the ground truth are attributable to the retrieved contexts. The formula for context recall is given as:

\begin{equation}
\text{Context Recall} = \frac{\lvert \text{GT attributed to retrieved contexts} \rvert}{\lvert \text{Total number of claims in GT} \rvert} \tag{1}
\end{equation}

\subsection{Overall Performance} 
\label{4.3} 

The performance of MQA models, shown in \hyperlink{overall}{Table 1}, demonstrates that ManuRAG\(_4\) consistently performs best across datasets. It achieves the highest factual correctness in MathQ, outperforming GPT-4O and GPT-4O with CoT, indicating its strength in handling mathematical reasoning. For MCQ tasks, it also scores highest in factual correctness, semantic similarity, and ROUGE (\hyperlink{rouge}{Table 2}). In RQ tasks, while the differences across models are smaller due to the simpler, text-based nature of the questions, ManuRAG\(_4\) still achieves the highest factual correctness, confirming its overall effectiveness.

MMRAG designs show varied performance across datasets. While ManuRAG\(_4\) consistently delivers the best results, other configurations, such as ManuRAG\(_1\) and ManuRAG\(_2\), exhibit strengths in specific tasks. For example, ManuRAG\(_1\) performs well on MCQ and RQ tasks by effectively leveraging image retrieval through links embedded in the textual context but performs poorly on MathQ, indicating limitations in its ability to handle mathematical reasoning with this retrieval method. The low performance of ManuRAG\(_3\) suggests that retrieving more images does not necessarily lead to better answers for MQA.

Across all datasets, the inclusion of CoT reasoning improves GPT-4O’s performance, showcasing its effectiveness in enhancing logical reasoning and precision for QA tasks. RAG\(_{hq}\) demonstrates better performance than RAG, highlighting the effectiveness of the multimodal data extraction method.

Semantic similarity scores are generally high across all models and datasets, reflecting strong alignment with reference answers. However, semantic similarity alone does not guarantee factual correctness, particularly in MathQ and MCQ, where precise reasoning and retrieval are critical. While semantic similarity remains useful for assessing general alignment, factual correctness is a more reliable indicator of performance in MQA tasks.

We include QA example comparisons across different models in \ref{appendix:QA}.

\hypertarget{overall}{}
\begin{table}[htbp]
\centering
\caption{Comparison of Different Metrics Across Methods for MQA (Best performance for each dataset is highlighted in \textbf{bold}).}

\begin{adjustbox}{width=0.8\linewidth} 
\small
\begin{tabular}{lcc}
\toprule
Category & Factual Correctness (\%) & Semantic Similarity (\%) \\
\midrule
\multicolumn{3}{c}{\textbf{MathQ}} \\
\midrule
GPT-4O & 49.91 & 87.03 \\
GPT-4O(CoT) & 56.22 & 86.85 \\
RAG & 59.51 & 87.66 \\
RAG$_{hq}$ & 59.48 & \textbf{87.90} \\
ManuRAG$_1$ & 19.14 & 77.03 \\
ManuRAG$_2$ & 56.69 & 87.32 \\
ManuRAG$_3$ & 55.31 & 87.32 \\
ManuRAG$_4$ & \textbf{62.03} & 87.82 \\
\midrule

\multicolumn{3}{c}{\textbf{MCQ}} \\
\midrule
GPT-4O & 6.99 & 77.60 \\
GPT-4O(CoT) & 78.58 & 94.79 \\
RAG & 85.22 & 96.52 \\
RAG$_{hq}$ & 86.38 & 96.62 \\
ManuRAG$_1$ & 86.82 & 95.27 \\
ManuRAG$_2$ & 60.54 & 91.20 \\
ManuRAG$_3$ & 61.62 & 90.15 \\
ManuRAG$_4$ & \textbf{87.61} & \textbf{96.82} \\
\midrule
\multicolumn{3}{c}{\textbf{RQ}} \\
\midrule
GPT-4O & 38.01 & 89.93 \\
GPT-4O(CoT) & 40.47 & 89.92 \\
RAG & 45.51 & 91.08 \\
RAG$_{hq}$ & 45.64 & 91.12 \\
ManuRAG$_1$ & 45.34 & 90.97 \\
ManuRAG$_2$ & 40.53 & \textbf{91.15} \\
ManuRAG$_3$ & 39.10 & 91.13 \\
ManuRAG$_4$ & \textbf{45.75} & 91.00 \\
\bottomrule
\end{tabular}
\end{adjustbox}
\end{table}

\hypertarget{rouge}{}
\begin{table}[htbp]
\centering
\caption{ROUGE Scores for MCQ (all) Across Different Models}
\small
\begin{tabular}{lc}
\toprule
\textbf{Model} & \textbf{ROUGE (\%)} \\
\midrule
GPT-4O & 10.35 \\
GPT-4O(CoT) & 77.61 \\
RAG & 86.10 \\
RAG$_{hq}$ & 86.71 \\
ManuRAG$_1$ & 81.19 \\
ManuRAG$_2$ & 63.13 \\
ManuRAG$_3$ & 58.88 \\
ManuRAG$_4$ & \textbf{86.77} \\
\bottomrule
\end{tabular}
\label{table:rouge_mcq}
\end{table}

\subsection{Context-based Performance}
\label{4.4}

While factual correctness evaluates the alignment of generated answers with GT, it does not assess the quality of retrieved context, which is critical in RAG-based models. Context precision and recall provide additional insights by measuring the relevance and coverage of the retrieved context, offering a deeper understanding of how well the model supports its generated answers. This section evaluates both the quality of the generated answers and the completeness of the retrieved context, as shown in \hyperlink{context}{Table 3}.

By aligning text and image data before embedding, ManuRAG\(_4\) achieves superior integration and consistent performance across tasks. It demonstrates robust performance across datasets, achieving the highest context precision for MathQ and MCQ, strong recall for MCQ and RQ, and effectively balancing relevance and coverage for retrieval-augmented tasks. RAG\(_{hq}\) excels in context precision for RQ, emphasizing the importance of high-quality indexed textual data, while standard RAG, with shorter indexed text, achieves the highest recall for MathQ and competitive recall for MCQ and RQ, highlighting a trade-off between precision and recall. In contrast, ManuRAG\(_1\), ManuRAG\(_2\), and ManuRAG\(_3\) show competitive precision for RQ but struggle with MathQ precision and lower recall for MathQ and MCQ, reflecting the limitations of separate vector stores for text and image data.

\hypertarget{context}{}
\begin{table}[htbp]
\centering
\caption{Context Recall and Precision for MathQ, MCQ, and RQ}
\label{table:context_recall_precision}
\begin{tabular}{lrrrrrr}
\toprule
{} & \multicolumn{3}{c}{Context Precision (\%)} & \multicolumn{3}{c}{Context Recall (\%)} \\
\cmidrule(lr){2-4} \cmidrule(lr){5-7}
Category & MCQ & MathQ & RQ & MCQ & MathQ & RQ \\
\midrule

RAG  & 81.95 & 92.56 & 92.24 & 33.39 & \textbf{67.83} & 81.23 \\
RAG$_{hq}$   & 85.35 & 93.20 & \textbf{93.74} & 34.61 & 57.00 & 81.27 \\
ManuRAG$_1$ & 79.41 & 33.66 & 92.79 & 33.36 & 30.95 & 81.47 \\
ManuRAG$_2$ & 80.89 & 33.33 & 92.52 & 32.01 & 22.98 & 80.63 \\
ManuRAG$_3$ & 79.41 & 33.33 & 92.65 & 33.44 & 20.47 & 80.86 \\
ManuRAG$_4$ & \textbf{86.84} & \textbf{93.53} & 91.84 & \textbf{37.88} & 62.26 & \textbf{81.56} \\
\bottomrule
\end{tabular}
\end{table}


\section{Conclusion And Future Work}
\label{sec:con}

Advances in digital manufacturing demand intelligent QA systems capable of integrating and analyzing complex multi-modal data. Traditional RAG methods often struggle to process such data effectively. ManuRAG addresses these challenges by introducing a novel MMRAG framework tailored for MQA. Experimental results demonstrate that ManuRAG$_4$ outperforms existing methods, offering valuable insights into selecting appropriate RAG strategies and aligning image and text data for diverse manufacturing scenarios. Beyond manufacturing, ManuRAG’s adaptable design positions it as a versatile tool for domain-specific QA in areas such as law, healthcare, and finance.

Despite its strengths, ManuRAG has limitations that present opportunities for improvements. First, adopting a stronger multi-modal embedding model, such as Jina-CLIP\cite{xiao2024jina}, could enhance the alignment of textual and visual information in the embedding space. Second, incorporating advanced retrieval designs from alternative RAG frameworks may further boost performance\cite{sharma2024og,10.1115/1.4067389}. Lastly, future iterations of ManuRAG should support additional data types, including CAD models\cite{wu2023cad}, videos\cite{huang2024vtimellm}, or time-series sensor data, to address the growing complexity and diversity of modern manufacturing information.

\appendix
\section{Prompt Design}
\label{appendix:Prompt}

\newtcolorbox{boxE}{
    enhanced, 
    boxrule = 0pt, 
    borderline = {0.75pt}{0pt}{main}, 
    borderline = {0.75pt}{2pt}{sub} 
}
The following shows the prompt designed for MathQ:
\begin{tcolorbox}[colback=violet!10!white, colframe=violet!85!black, title=MathQ Prompt]
As a manufacturing expert, solve the math question step by step using the provided context and your knowledge of manufacturing calculations. Make sure to clearly explain your reasoning:

\texttt{\{context\_str\}}

Question:

Query: \texttt{\{query\_str\}}

Answer:
\end{tcolorbox}

The following presents the prompt designed for MCQ:
\begin{tcolorbox}[colback=violet!10!white, colframe=violet!85!black, title=MCQ Prompt]
As a manufacturing expert, answer the following multi-choice questions based on the provided context step by step. Select the correct letter choices only.

Answer format:

"Explanation": "Your explanation here.", "YourChoice": "Answer. (d), (e), and (g)."

\texttt{\{context\_str\}}

Answer the following question:

Query: \texttt{\{query\_str\}}

Answer:
\end{tcolorbox}

The following shows the prompt designed for RQ:
\begin{tcolorbox}[colback=violet!10!white, colframe=violet!85!black, title=RQ Prompt]
You are an expert in manufacturing major. You need to answer the following short response question based on the context and your knowledge of manufacturing processes and materials. Please provide a detailed explanation and justify your answer. Be concise, clear, and accurate.

\texttt{\{context\_str\}}

Answer the following question:

Query: \texttt{\{query\_str\}}

Answer:
\end{tcolorbox}

\section{QA Examples}
\label{appendix:QA}

This section presents QA examples of different types and their answers generated by three different models, specifically comparing GPT-4O (CoT), traditional RAG, and the best-performing ManuRAG variant.

\subsection{MathQ Example}
The following is an example of QA pairs from MathQ. ManuRAG$_4$ generates correct answers, whereas RAG and GPT-4O (CoT) fail. GPT-4O (CoT) cannot reference Table 4.1 or the associated formulas from the manufacturing documents, limiting its accuracy. While RAG retrieves relevant formulas from the text data, it is constrained by its inability to access visual information, such as the data in Table 4.1 (from the textbook).

\begin{tcolorbox}[colback=blue!5!white,colframe=blue!75!black,title=MathQ Problem]
The starting diameter of a shaft is 25.00 mm. This shaft is to be inserted into a hole in an expansion fit assembly operation. To be readily inserted, the shaft must be reduced in diameter by cooling. Determine the temperature to which the shaft must be reduced from room temperature (20℃) to reduce its diameter to 24.98 mm. Refer to Table 4.1.
\end{tcolorbox}

\begin{tcolorbox}[colback=green!5!white,colframe=green!75!black,title=Correct Answer]
For steel, \(\alpha = 12 \times 10^{-6} \, \text{mm/mm/℃}\) according to Table 4.1.  
Revise Eq. (4.1) to:
\[
D_2 - D_1 = \alpha D_1 (T_2 - T_1)
\]
Substitute the known values:
\[
24.98 - 25.00 = 12 \times 10^{-6} \cdot 25.00 \cdot (T_2 - 20)
\]
Simplify:
\[
-0.02 = 300 \times 10^{-6} \cdot (T_2 - 20)
\]
\[
-0.02 = 0.0003(T_2 - 20)
\]
\[
-0.02 = 0.0003T_2 - 0.006
\]
\[
-0.02 + 0.006 = 0.0003T_2
\]
\[
-0.014 = 0.0003T_2
\]
\[
T_2 = \frac{-0.014}{0.0003}
\]
\[
T_2 = \underline{-46.67^\circ \text{C}}
\]
\end{tcolorbox}

\begin{tcolorbox}[colback=yellow!10!white,colframe=yellow!70!black,title=GPT-4O (COT)]
To solve this problem, we need...

\textbf{Calculation:}  
Step 1: Understand the Concept of Thermal Contraction... 

Step 4: Use the Provided Table to Find the Coefficient of Linear Expansion...

Since specific data \underline{(Table 4.1) is mentioned but not provided here}, let's assume...

Step 6: Calculate the Final Temperature...

Using the assumed coefficient of linear expansion, the calculated temperature is:
\[
T_2 = \underline{-52.73^\circ \text{C}}
\]

\textbf{Conclusion:}  
The shaft must be cooled to approximately \underline{\(-52.73^\circ \text{C}\)} to achieve the desired diameter reduction from 25.00 mm to 24.98 mm.
\end{tcolorbox}

\begin{tcolorbox}[colback=gray!10!white,colframe=gray!70!black,title=RAG]
To solve this problem, we need...

\textbf{Calculation}  
The formula for the change in diameter due to thermal expansion or contraction is given by:
\[
D_2 - D_1 = \alpha \cdot D_1 \cdot (T_2 - T_1)
\]
First, calculate the change in diameter...
Using the coefficient of linear thermal expansion (\(\alpha\)), the temperature is calculated as:
\[
T_2 = \underline{-48.39^\circ \text{C}}
\]

\textbf{Conclusion:}  
The shaft must be cooled to approximately \underline{\(-48.39^\circ \text{C}\)} to achieve the desired diameter reduction for the expansion fit assembly operation.
\end{tcolorbox}

\begin{tcolorbox}[colback=cyan!5!white,colframe=cyan!75!black,title=ManuRAG$_4$]
To solve this problem, we need...

The change in diameter due to thermal contraction is given by:
\[
D_2 - D_1 = \alpha \cdot D_1 \cdot (T_2 - T_1)
\]

\textbf{Calculation:}  
1. Calculate the change in diameter: …  
2. Substitute the values into the formula: …  
3. Calculate the temperature change:
\[
T_2 = \frac{-0.02}{0.0003} + 20
\]
\[
T_2 = \underline{-46.67^\circ \text{C}}
\]

\textbf{Conclusion:}  
The shaft must be cooled to approximately \underline{\(-46.67^\circ \text{C}\)} to reduce its diameter from 25.00 mm to 24.98 mm for the expansion fit assembly operation.
\end{tcolorbox}

\subsection{MCQ Example}
This question represents a MCQ problem for the manufacturing materials. GPT-4O (CoT) fails to generate a completely correct answer, whereas both RAG and ManuRAG$_4$ provide accurate responses. RAG and ManuRAG4 achieves the correct result by effectively utilizing retrieved textual context from the manufacturing textbook.

\begin{tcolorbox}[colback=blue!5!white,colframe=blue!75!black,title=MCQ Problem]
Which of the following materials are used as fibers in fiber-reinforced plastics (FRPs) (select more than one):  
(a) aluminum oxide,  
(b) boron,  
(c) carbon/graphite,  
(d) epoxy,  
(e) Kevlar 49,  
(f) S-glass, and  
(g) unsaturated polyester.
\end{tcolorbox}

\begin{tcolorbox}[colback=green!5!white,colframe=green!75!black,title=Correct Answer]
The correct materials used as fibers in FRPs are:  
\underline{\textbf{(a)}, \textbf{(b)}, \textbf{(c)}, \textbf{(e)}, and \textbf{(f)}}.
\end{tcolorbox}

\begin{tcolorbox}[colback=yellow!10!white,colframe=yellow!70!black,title=GPT-4O (CoT)]
\textbf{Explanation:}  
Fiber-reinforced plastics (FRPs) are composite materials made of a polymer matrix reinforced with fibers to enhance their mechanical properties. In the given options:  
- \textbf{(a)} Aluminum oxide is typically not used as a fiber in FRPs; rather, it's more common as a ceramic or in powder form for reinforcement.  
- \textbf{(b)} Boron is indeed used as a fiber in certain high-performance FRPs.  
- \textbf{(c)} Carbon/graphite is one of the most common and effective fiber reinforcements.  
- \textbf{(d)} Epoxy is generally a type of polymer matrix in FRPs, not a fiber.  
- \textbf{(e)} Kevlar 49 is an aramid fiber known for high strength and is used in FRPs.  
- \textbf{(f)} S-glass is a specialized type of glass fiber used in FRPs for higher strength.  
- \textbf{(g)} Unsaturated polyester is also a type of polymer matrix and not used as a fiber.  

\textbf{Your Choice:} \underline{\textbf{(b)}, \textbf{(c)}, \textbf{(e)}, and \textbf{(f)}}.
\end{tcolorbox}

\begin{tcolorbox}[colback=gray!10!white,colframe=gray!70!black,title=RAG]
\textbf{Explanation:}  
Fiber-reinforced plastics (FRPs) are composite materials that consist of a polymer matrix reinforced with fibers. \underline{According to the provided context}, the principal fiber materials used in FRPs include glass (such as E-glass and S-glass), carbon, and Kevlar 49. 
Additionally, less common fibers mentioned are boron, silicon carbide (SiC), aluminum oxide, and steel. Epoxy and unsaturated polyester are not fibers; they are types of polymer matrices used in FRPs.  

\textbf{Your Choice:} \underline{\textbf{(a)}, \textbf{(b)}, \textbf{(c)}, \textbf{(e)}, and \textbf{(f)}}.
\end{tcolorbox}

\begin{tcolorbox}[colback=cyan!5!white,colframe=cyan!75!black,title=ManuRAG$_4$]
\textbf{Explanation:}  
The question asks for materials used as fibers in fiber-reinforced plastics (FRPs). \underline{Based on the provided context}, the materials used as fibers in FRPs include:  
- \textbf{(a)} Aluminum oxide (\( \text{Al}_2\text{O}_3 \)) is mentioned as a less common fiber material in FRPs.  
- \textbf{(b)} Boron is mentioned as a fiber used in advanced composites.  
- \textbf{(c)} Carbon/graphite is a principal fiber material in FRPs.  
- \textbf{(e)} Kevlar 49 is listed as a principal fiber material in FRPs.  
- \textbf{(f)} S-glass is a type of glass fiber used in FRPs.  

Epoxy and unsaturated polyester are not fibers; they are matrix materials used in FRPs.  

\textbf{Your Choice:} \underline{\textbf{(a)}, \textbf{(b)}, \textbf{(c)}, \textbf{(e)}, and \textbf{(f)}}.
\end{tcolorbox}

\subsection{RQ Example}
The following is an example of review question in the dataset. GPT-4O (CoT) provides a comprehensive list but includes less relevant variables such as pressure and barriers, leading to less focus. RAG identifies the key variables accurately, leveraging retrieved context but lacks consistency in prioritizing variables. ManuRAG$_4$ achieves the most concise and correct explanation by effectively retrieving and integrating context from relevant sources.
\begin{tcolorbox}[colback=blue!5!white,colframe=blue!75!black,title=RQ Problem]
What are the important variables that affect mass diffusion?
\end{tcolorbox}

\begin{tcolorbox}[colback=green!5!white,colframe=green!75!black,title=Correct Answer]
According to Fick's first law, mass diffusion depends on the following variables:
\begin{itemize}
    \item \textbf{Diffusion Coefficient (\(D\))}: The diffusion coefficient rises rapidly with temperature, so temperature is an important variable.
    \item \textbf{Concentration Gradient (\(dc/dx\))}: The difference in concentration of diffusing species across a distance.
    \item \textbf{Contact Area (\(A\))}: The area through which diffusion occurs.
    \item \textbf{Time (\(t\))}: The duration of the diffusion process.
\end{itemize}
\end{tcolorbox}

\begin{tcolorbox}[colback=yellow!10!white,colframe=yellow!70!black,title=GPT-4O (CoT)]
\textbf{Explanation:}  
Mass diffusion is a critical process in manufacturing and materials science. The variables that affect diffusion include:
\begin{enumerate}
    \item \textbf{Temperature}: Temperature is perhaps the most influential factor affecting diffusion. Higher temperatures increase atomic motion, accelerating diffusion.
    \item \textbf{Concentration Gradient}: The concentration gradient, or the difference in concentration of diffusing species between two points, directly affects diffusion rates.
    \item \textbf{Diffusing Species}: The size, mass, and chemical nature of the particles being diffused influence diffusion rates.
    \item \textbf{Medium or Matrix Structure}: The structure of the material through which diffusion occurs is critical.
    \item \textbf{Pressure}: In gases and, to some extent, in liquids, pressure can influence diffusion rates.
    \item \textbf{Time}: Over longer periods, the extent of diffusion naturally increases as particles spread through the medium.
    \item \textbf{Presence of Barriers or Interfaces}: Physical barriers or interfaces, such as coatings or phase boundaries, can alter the pathway and rate of diffusion.
\end{enumerate}
\end{tcolorbox}

\begin{tcolorbox}[colback=gray!10!white,colframe=gray!70!black,title=RAG]
\textbf{Explanation:}  
Mass diffusion in solids, particularly metals, is influenced by several key variables:
\begin{enumerate}
    \item \textbf{Temperature}: A critical factor affecting diffusion, as higher temperatures increase atomic mobility.
    \item \textbf{Concentration Gradient (\(dc/dx\))}: Represents the difference in concentration of diffusing species across a distance.
    \item \textbf{Diffusion Coefficient (\(D\))}: A material-specific parameter that quantifies how easily atoms can move through a material.
    \item \textbf{Crystal Structure and Defects}: Vacancies and other imperfections in the crystal structure facilitate atomic motion.
    \item \textbf{Material Properties}: Atomic size, bonding, and lattice structure influence diffusion behavior.
    \item \textbf{Surface Area (\(A\))}: The area of the boundary through which diffusion occurs.
\end{enumerate}
\end{tcolorbox}

\begin{tcolorbox}[colback=cyan!5!white,colframe=cyan!75!black,title=ManuRAG$_4$]
\textbf{Explanation:}  
The important variables that affect mass diffusion include:
\begin{enumerate}
    \item \textbf{Temperature}: A critical factor in diffusion. Higher temperatures increase thermal agitation, allowing atoms to move more freely.
    \item \textbf{Concentration Gradient (\(dc/dx\))}: The driving force for diffusion, representing the difference in concentration of atoms across a distance.
    \item \textbf{Diffusion Coefficient (\(D\))}: A material-specific parameter that quantifies the ease with which atoms can move through a material.
    \item \textbf{Time (\(t\))}: The duration over which diffusion occurs affects the extent of diffusion.
    \item \textbf{Area of the Boundary (\(A\))}: The area through which diffusion occurs impacts the rate of mass transfer.
\end{enumerate}
\end{tcolorbox}

\bibliographystyle{unsrtnat} 
\bibliography{cas-refs} 





\end{document}